# Multi-instrument Comparative Study of Temperature, Number Density, and Emission Measure during the Precursor Phase of a Solar Flare

Nian Liu[1], Ju Jing[1,2], Yan Xu[1,2], and Haimin Wang[1,2]
[1] Institute for Space Weather Sciences, New Jersey Institute of Technology, Newark, NJ 07102-1982, USA; nl244@njit.edu
[2] Big Bear Solar Observatory, New Jersey Institute of Technology, 40386 North Shore Lane, Big Bear City, CA 92314-9672, USA

## Abstract

We present a multi-instrument study of the two precursor brightenings prior to the M6.5 flare (SOL2015-06-22T18:23) in the NOAA Active Region 12371, with a focus on the temperature ($T$), electron number density ($n$), and emission measure (EM). The data used in this study were obtained from four instruments with a variety of wavelengths, i.e., the Solar Dynamics Observatory's Atmospheric Imaging Assembly (AIA), in six extreme ultraviolet (EUV) passbands; the Expanded Owens Valley Solar Array (EOVSA) in microwave (MW); the Reuven Ramaty High Energy Solar Spectroscopic Imager (RHESSI) in hard X-rays (HXR); and the Geostationary Operational Environmental Satellite (GOES) in soft X-rays (SXR). We compare the temporal variations of $T$, $n$, and EM derived from the different data sets. Here are the key results. (1) GOES SXR and AIA EUV have almost identical EM variations ($1.5$–$3 \times 10^{48}$ cm$^{-3}$) and very similar $T$ variations, from 8 to 15 million Kelvin (MK). (2) Listed from highest to lowest, EOVSA MW provides the highest temperature variations (15–60 MK), followed by RHESSI HXR (10–24 MK), then GOES SXR and AIA EUV (8–15 MK). (3) The EM variation from the RHESSI HXR measurements is always less than the values from AIA EUV and GOES SXR by at most 20 times. The number density variation from EOVSA MW is greater than the value from AIA EUV by at most 100 times. The results quantitatively describe the differences in the thermal parameters at the precursor phase, as measured by different instruments operating at different wavelength regimes and for different emission mechanisms.

*Unified Astronomy Thesaurus concepts:* Solar flares (1496); Solar x-ray flares (1816); Solar radio flares (1342); Solar physics (1476); Solar particle emission (1517); The Sun (1693)

## 1. Introduction

Flare precursor brightenings, shown as small-scale emissions in various wavelengths, such as optical, ultraviolet/extreme ultraviolet (EUV), soft X-ray (SXR), hard X-ray (HXR), and microwave (MW), have been observed prior to many flares and regarded as a result of localized magnetic reconnection and the subsequent small-scale energy release (see, e.g., the reviews by Martin 1980 and Hoven & Hurford 1986, and the statistical study by Gyenge et al. 2016). Most of the energy released in the precursor phase is considered to be thermal (Awasthi et al. 2014), observed in multiwavelength. More than half of the precursors are visible in MW emissions 60 minutes before the associated SXR flare emissions reach their maxima (Fernandes et al. 2016). Some changes in the magnetic characteristics of sunspot groups may even occur two days before the main flares (Abramov-Maximov et al. 2015). Most of the precursors occur within a distance of about 0.1 diameters of the sunspot group from the site of the major flare (Gyenge et al. 2016).

Although the concept of flare precursors was initially introduced almost 60 yr ago (Bumba & Křivský 1959), and precursor brightenings have been observed in many different wavelengths, such small-scale emissions have thus far not been well understood and characterized. Despite the growing literature of precursor studies, especially after the new millennium, a comprehensive, comparative study of precursors in multiwavelength is particularly lacking. For the same event, different data sets may come to inconsistent conclusions. For instance, before an X-class flare, a significant precursor was found in the HXR emission, but was completely absent in the MW observation (Zimovets et al. 2009). Such a difference, and the underlying physics, have not been addressed in previous studies.

Some of the difference in the results is due to the varying definitions of precursors. For example, some studies attribute B-/C-class flares prior to a major flare to precursors, while for others, any brightness enhancement observed in the vicinity of and prior to the main flare is considered a precursor, regardless of the wavelengths. The time intervals for these kinds of precursor observations are also very different, ranging from 25 minutes (Hernandez-Perez et al. 2019) to 10 hr (Sterling et al. 2011). Other studies consider very long periodic pulsations (Tan et al. 2016; Li et al. 2020a) and quasiperiodic pulsations (Chen et al. 2019; Li et al. 2020b) during the preflare phase to be precursors. Such kinds of precursors usually have periods in the unit of minutes.

In this study, a flare precursor is defined as an increase in brightness, identified with the EUV hot channels (94 Å and 131 Å) of the Atmospheric Imaging Assembly (AIA; Lemen et al. 2012) on board the Solar Dynamics Observatory (SDO; Pesnell et al. 2012), prior to a flare. Presumably, precursors may be related to two types of small-scale magnetic structures near the magnetic polarity inversion line (PIL), as demonstrated by the magnetohydrodynamics (MHD) simulation by Kusano et al. (2012). These two types of magnetic structures are reversed shear (RS) and opposite polarity (OP), both of which have been tested for their applicability by MHD simulations (Bamba & Kusano 2018) and observed before major flares. For example, an RS-type magnetic structure was seen before an X-class flare (Bamba et al. 2017). Taking advantage of the high-resolution observation of the 1.6 m Goode Solar Telescope (GST) at Big Bear Solar Observatory, a magnetic channel structure (seen as an elongated alternating structure of positive and negative polarities) was recognized as an OP-type structure and was found to be associated with the precursor brightenings before an M6.5 flare (Wang et al. 2017).

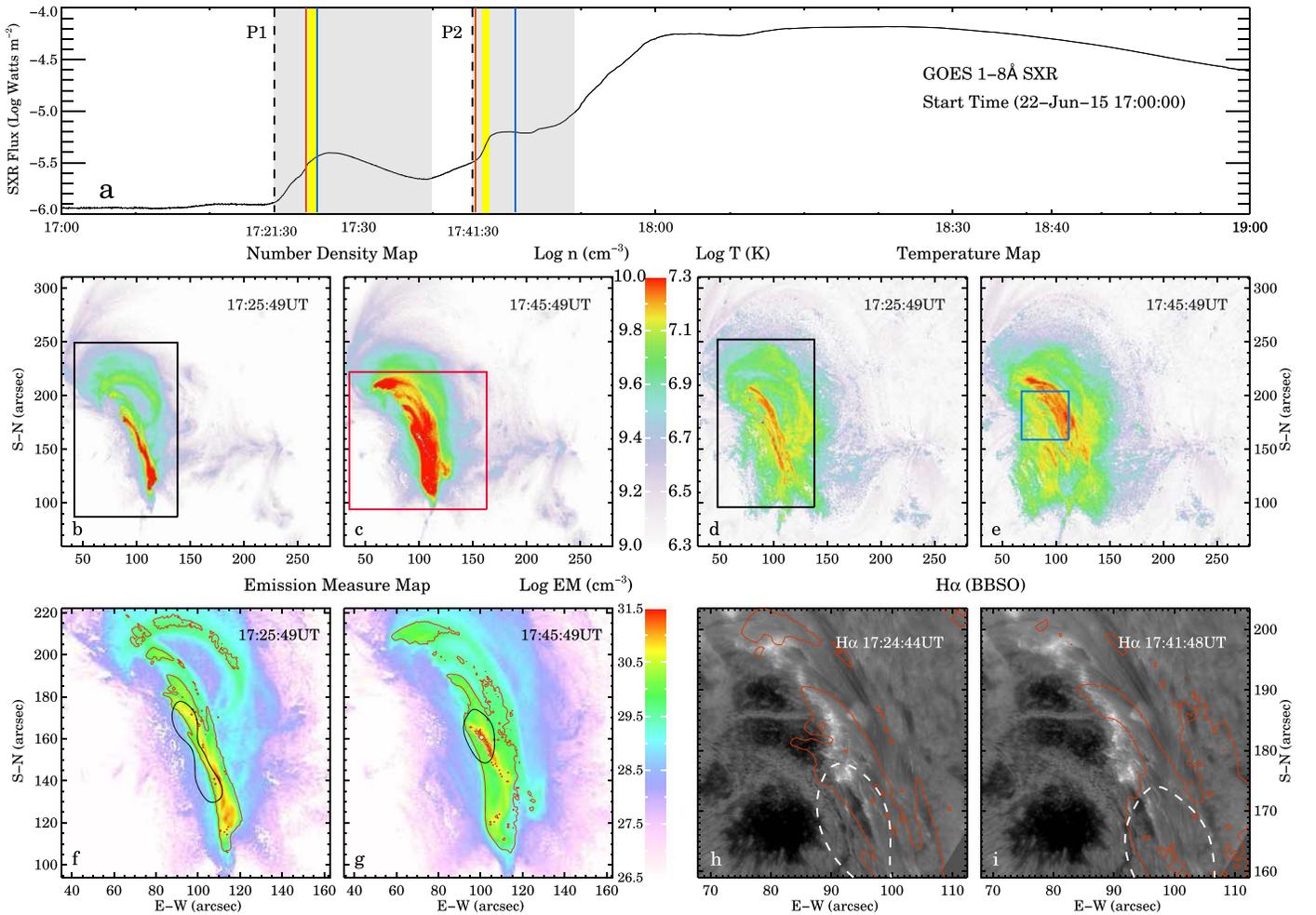

**Figure 1.** (a) GOES SXR flux light curve, with the gray shaded areas P1 and P2 denoting the two precursor periods. The red and blue lines mark the times of the $n$ and $T$ maps in (b)–(e), respectively. The yellow shaded areas indicate the times of the HXR imaging integration times in (f)–(g). (b)–(c) $n$ maps, derived from AIA DEM analysis, at the selected times of P1 and P2. (d)–(e) $T$ maps, derived from AIA DEM analysis, at the selected times of P1 and P2. The black rectangular boxes in panels (b) and (d) are drawn to define the areas used in the calculation of the average $n$ and $T$ (shown in Figures 6 and 4). The red square box in panel (c) is drawn to mark the FOV of panels (f)–(g), and the small blue square box in (e) indicates the FOV of (h)–(i). (f)–(g) Close-up views of the DEM maps of $\log T = [6.85, 7.35]$, superimposed with red contours of 30% of the density maximum and black contours of 80% of the RHESSI HXR intensity maximum in 6–12 KeV. The HXR imaging time ranges of the two precursors are 17:25:00–17:26:00 UT and 17:42:38–17:43:20 UT, respectively. (h)–(i) Two snapshots of GST H$\alpha$+0.6 Å images showing the two precursor brightenings. The red contours show 30% of the density maximum, the same as in (f)–(g). The white dashed contours show 80% of the RHESSI HXR intensity maximum, the same as the black contours in (f)–(g).

Here, we study the emission properties of the same two precursors. The flare starts at 17:39 UT, peaks at 18:23 UT, and ends at 18:51 UT, according to the NOAA Space Weather Prediction Center (SWPC) flare report. However, the Geostationary Operational Environmental Satellite (GOES) SXR light curve clearly displays two flux enhancements at ∼17:24 UT and ∼17:42 UT (denoted as P1 and P2 in Figure 1a, respectively), preceding the impulsive phase of the flare. The first one is listed as a C3.9 flare in the NOAA SWPC report. Meanwhile, both chromospheric H$\alpha$ and coronal EUV AIA observations show brightenings in the close vicinity of the magnetic PIL, which spatially coincide with the main flare afterwards. Therefore, the two brightenings at ∼17:24 UT and ∼17:42 UT are regarded as the precursors of the M6.5 flare at 18:23 UT (Wang et al. 2017). The high-resolution flows during the pre-eruption phase and their relation to magnetic field changes have been presented in previous studies (Li et al. 2017; Wang et al. 2018). These precursors signify near-surface magnetic reconnection in the early stage of the flare, and may contribute to the onset of the main flare (Wang et al. 2017).

In this paper, we focus on the thermal behaviors of the precursors in the hour before the flare. These two precursors show simple emission structures that are mostly confined to a small area at the magnetic PIL (see Figure 2). In general, energy in the form of nonthermal emissions is released in HXR and MW during the solar flares. The dominant emission mechanism of HXR is bremsstrahlung, with electrons precipitating at the footpoints and loop top of a magnetic flux-rope structure, which can be approximated by a thick-target model (Brown 1971). On the other hand, MWs are emitted by gyrosynchrotron (nonthermal) emission and free–free bremsstrahlung (thermal) emission mechanisms (Aschwanden 2002).

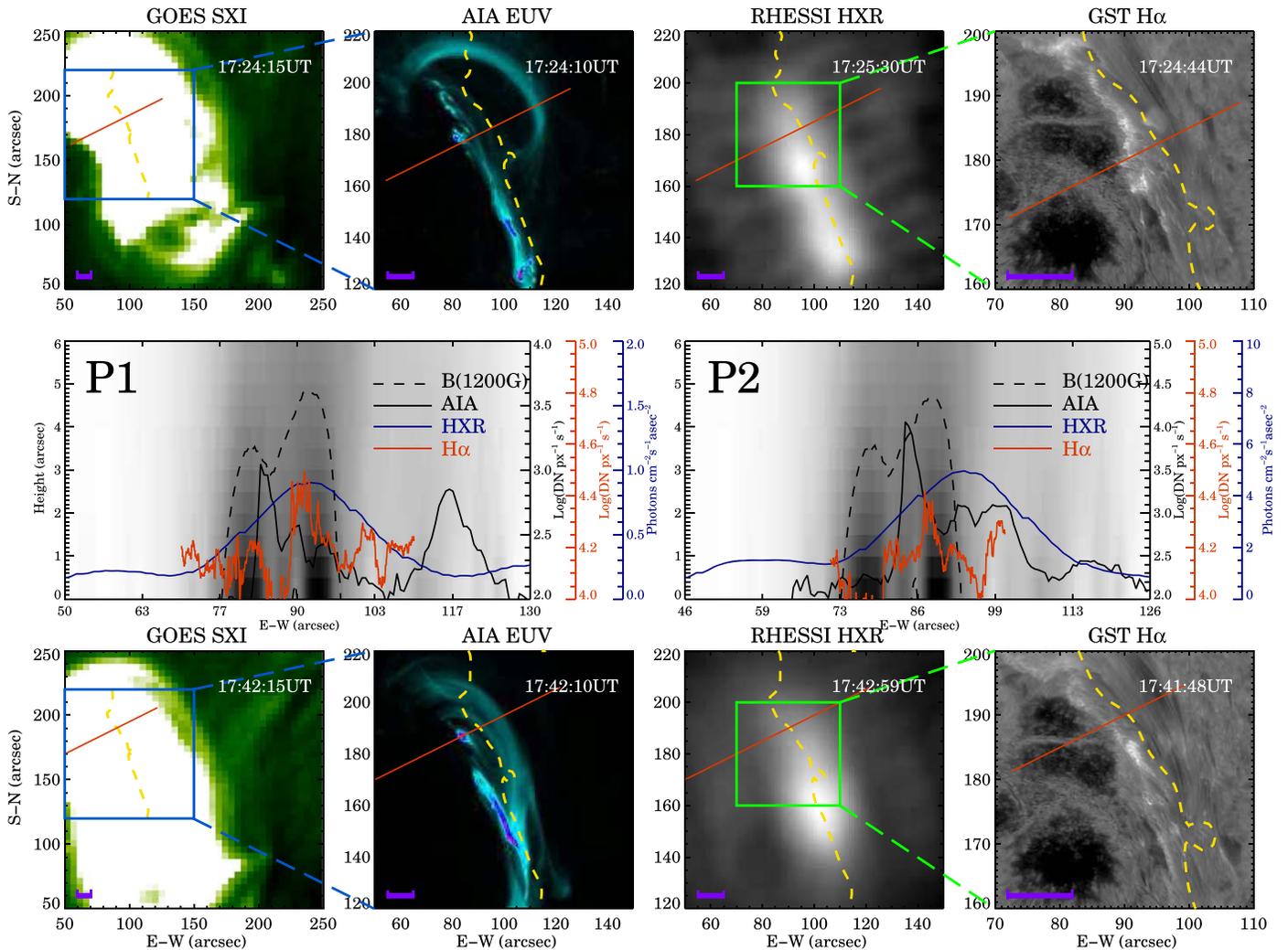

**Figure 2.** The top and bottom panels show the precursor emission maps from four wavelengths (GOES/SXR, AIA 131 Å, RHESSI HXR, and GST Hα) at the two precursor times P1 and P2, respectively. The blue boxes in the GOES SXI panels mark the FOV of the AIA EUV and RHESSI HXR images. The green boxes in the RHESSI HXR images mark the FOV of the GST Hα images. The yellow dashed lines in each image mark the locations of the PILs of the corresponding precursors, with the PIL being defined by the zero-value contour of the vertical magnetic field obtained by SDO/HMI. For each precursor, the slits are centered on the same location, but the lengths of the slits are not necessarily equal. The purple segments in each of the top and bottom panels indicate the estimated D values. The middle panel shows the photometric intensity profiles along the slit and the spatial distribution of the magnetic field strength calculated from the extrapolated 3D NLFFF. The dashed line marks the magnetic field of 1200 G.

During this process, the thermalization of the precipitated nonthermal electrons results in the formation of hot dense plasma, which evaporates into the corona and leads to the EUV, HXR, and SXR emissions via thermal bremsstrahlung (Abramov-Maximov et al. 2002; Carruthers 2003; Hudson 2011). The emission mechanisms at HXR, MW, SXR, and EUV during the precursor phase of a flare are not different from those during the main flare. The multi-instrument study of the thermal emissions will provide a meaningful comparison of different wavelengths over a large temperature range.

## 2. Data Sets and Methods

The thermal parameters analyzed in this paper include temperature ($T$), the electron number density ($n$) of the emitting plasma, and the emission measure (EM), which can be obtained from a variety of data sets with different methodologies. For example, $T$ can be directly obtained from GOES SXR flux measurements, using the formulae for the reduction of GOES X-ray data (Thomas & Crannell 1985). In their definition, $T$ is the temperature of an isothermal plasma that produces the same ratio of responses in two wavelength intervals (0.5–4 Å and 1–8 Å) as observed. $T$ can also be inferred from the EUV observations of AIA/SDO with differential emission measure (DEM) inversion techniques, e.g., the sparse inversion method by Cheung et al. (2015). Moreover, the X-ray spectral analysis of Reuven Ramaty High Energy Solar Spectroscopic Imager (RHESSI) data using the Object Spectral Executive (Schwartz et al. 2002) interface provides another way to derive $T$. By doing the simplest photon flux fitting with combined Variable Thermal Function (*vth*) and a Thick-target Nonthermal Function (*thick2_vnorm*), both $T$ and EM can be obtained. $T$ and $n$ can also be derived from Expanded Owens Valley Solar Array (EOVSA) MW data with a combined gyro- and free-to-free emission fitting method (Fleishman & Kuznetsov 2010).

The calculations of $T$, $n$, and EM with data from different instruments are introduced in detail in the following subsections.

### 2.1. AIA

The DEM analysis was performed with level 1.5 data of six AIA/EUV passbands (94, 131, 171, 193, 211, and 335 Å), with a spatial resolution of 0″.6 per pixel. Using the sparse inversion method of Cheung et al. (2015), the DEM solutions were obtained in the course of the precursors (17–18 UT) at a cadence of 48 s. DEM maps were obtained from inversion on a pixel-by-pixel basis on a temperature grid with $\log T = 5.7, 5.8, 5.9 \ldots 7.6, 7.7$ (Jing et al. 2017). To compare with the GOES and RHESSI results in this study, the EM of AIA is defined as the volume-integrated total emission measure in the units of $\mathrm{cm}^{-3}$.

At each pixel $i$, $T$ and EM are calculated as:

$$T_i = \frac{\sum_t T_{t,i} \cdot \mathrm{DEM}_{t,i}}{\sum_t \mathrm{DEM}_{t,i}}; \tag{1}$$

$$\mathrm{EM}_i = A_i \cdot \sum_t \mathrm{DEM}_{t,i}, \tag{2}$$

where the subscripts $t$ and $i$ represent each temperature grid and each pixel, respectively, and $A_i$ is the area of the spatial sampling.

At each pixel $i$, $n$ is defined as:

$$n_i = \sqrt{\frac{\sum_t \mathrm{DEM}_{t,i}}{L}}, \tag{3}$$

where $L$ is the length of plasma along the line of sight that contributes to the emission over the brightening strip. The brightening strip is a slender area that displays clear emissions in multiwavelengths, including AIA EUV and Hα, during the precursor phase. Since $L$ is not accessible from observation, the value of $L$ is represented by the emission width ($D$). We assume that the emissions are isotropic, thus the cross-sectional width of the plasma measured in the x–y plane could represent the plasma length along the line of sight (i.e., $L \sim D$). The detailed steps are as follows. At each time, we performed 1D Gaussian fittings to the temperature distribution along a series of cross-sectional cuts (the red lines in each panel of Figure 2 constitute one example). The Gaussian FWHM for each fitting is a simple representation of the cross-sectional width of the plasma, and its average, weighted by its fitted peak DEM value, was defined as $D$, which was used to estimate the length of the plasma along the line of sight. It was found in this case that the value of $D$ lies in a range of 5″–10″. AIA DEM uncertainties are estimated using the data-to-noise ratio,[3] where the uncertainty of length of the plasma ($L$) is estimated as $\sigma$ in the Gaussian fitting.

### 2.2. GOES

$T$ and EM can also be computed from GOES SXR measurements. The essential ideas underlying the computation have been described by Thomas & Crannell (1985; see Equations (1)–(8)). The X-ray values, $B_i$, depend on EM and the detector responses, $b_i$:

$$B_i = \mathrm{EM} \cdot b_i(T) \tag{4}$$

($i = 4$ denotes the 0.5–4 Å detector and $i = 8$ the 1–8 Å detector). Briefly speaking, assuming the entire plasma volume to be isothermal, the temperature equals the theoretical value that could produce the same ratio of response of two detectors:

$$R(T) = B_4/B_8 = b_4(T)/b_8(T). \tag{5}$$

The ratio $R$ is only a function of $T$. The EM can then be calculated simultaneously by taking the X-ray flux measurements and the detector's response ratio into account for a certain temperature:

$$\mathrm{EM} = B_4/b_4(T) = B_8/b_8(T). \tag{6}$$

These formulae were updated by White & Thomas (2005) in order to modify the temperature measurements in hot flare (∼35 MK) conditions. The advantage of this method is that only a simple analytic curve fitting is required, and hence the uncertainties are small over a vast temperature range (within 2% of the temperature and 5% of EM between 5 and 30 million degrees, in accordance with the expressions for determining $T$ and EM from the GOES measurements (Thomas & Crannell 1985)). The disadvantage, on the other hand, is that its accuracy will be diminished if more than one active region (AR) is present on the solar disk. In fact, there were four ARs on the solar disk during this precursor period. AR 12371 was the only AR near the center of the solar disk, while the other three were very close to the west limb. Except for this flare, no other flares have been found within 10 hr before and after this one, according to the NOAA SWPC report. Therefore, we assume that the flare emission in AR 12371 is dominant in the GOES SXR measurements.

### 2.3. RHESSI

RHESSI (Lin et al. 2002, 2003) measures photon flux in HXR and gamma-ray, produced by high-energy electrons via bremsstrahlung. The best pixel resolution of RHESSI synthesis maps is about 2″.3 (Lin 2002). In principle, a 2 s time interval (half a rotation of the rotational modulation collimators carried on RHESSI) provides enough Fourier components for image reconstruction. This can be changed according to the photon count rate in practice. The CLEAN algorithm (Hurford et al. 2003) was used in the HXR imaging with front detectors 3–8. The total field of view (FOV) was set to 128″ × 128″, with a pixel size of 1″. The spectral resolution varies at different energy ranges, from 1 keV resolution at 3 keV level to 5 keV resolution at 5 MeV level. In this study, we chose an empirical binning code (#14) provided by the RHESSI GUI. The spectra in different time intervals can be fitted with various emission models. In this study, the HXR spectra are fitted using the Variable Thermal Model (*vth*) and the second version of the Thick-target Bremsstrahlung with Independent Normalization Model (*thick2_vnorm*). Meanwhile, *albedo* is included for correcting for albedo, and *pileup_mod* is also applied to add pileup effects to the models. The backgrounds of the HXR spectral fitting were selected carefully. For energy ranges of 6–25 keV, we took the background before the first precursor and considered it to be constant during the flare. However, for energy ranges greater than 25 keV, the HXR emissions of precursors are not so significant compared to the fluctuation of the background. The emissions due to precursors usually appear as several clear but transient spikes in the HXR flux profile, and they should be separated from the changing background. In this case, the background time was selected over the whole flaring time, except for the spikes, and the interpolated count rates were used as the background. The physical parameters of the thermal components, such as EM and $T$, are obtained from the *vth* model. The uncertainties of RHESSI are estimated using a Chi-square ($\chi^2$) test in spectra fittings. In the process of fitting the HXR spectrum, we noticed that the thermal component (*vth*) dominates under 20 keV in the vast majority of the fitting time intervals; specifically, in almost all cases of the first precursor and more than half the cases of the second precursor. During the precursor phase, however, the maximum flux counts of the nonthermal component continuously increased, and, finally, the nonthermal emission peaked at 17:58 UT (Wang et al. 2017), which is 15 minutes after the second precursor. Two examples of the fitting results of the RHESSI photon counts and normalized residuals are shown in Figure 3.

---
[3] https:www.lmsal.com/c̃heung/AIA/tutorial_dem/sparse_exercise1.pro

### 2.4. EOVSA

The MW data comes from EOVSA (Hurford et al. 1984; Gary & Hurford 1994; Gary et al. 2018). EOVSA is a solar-dedicated MW imaging array operating in the frequency range of 1–18 GHz. In this study, the sequential spectral fit (Gary et al. 2013) is performed by combining gyrosynchrotron emission and free–free emission, which are responsible for the lower and higher frequencies, respectively. The fast gyrosynchrotron codes (Fleishman & Kuznetsov 2010; Fleishman et al. 2011; Gary et al. 2013) were applied to quantitatively define the gyrosynchrotron source function, which is designed for both isotropic and anisotropic electron distributions. The uncertainties are estimated using a Chi-square ($\chi^2$) test in spectra fittings. An important assumption of a uniform emission source is made to perform the sequential spectral fit. Although the broadband MW spectrum during the main flare indicates the source is spatially nonuniform, at the precursor phase, the spectral line was fitted with a quasi-uniform source because reasonably narrow spectra are observed during the precursor phase (Wang et al. 2017). The size of the emission source is estimated as having a depth of $10''$ and an area of $10'' \times 30''$, based on the observation of the GST H$\alpha$ image. By performing such a fitting for a thermal source, three free thermal parameters, including $n$, $T$, and magnetic field strength ($B$), are derived (Nita et al. 2015). As a follow-up study to Wang et al. (2017), this study uses the same EOVSA MW data analysis results. The detailed methodology is described in Fleishman et al. (2015).

### 2.5. Magnetic Field Extrapolation

The 3D coronal magnetic field was reconstructed by performing nonlinear force-free field (NLFFF) extrapolation with the weighted optimization method (Wiegelmann 2004). The *hmi.sharp_cea_720s* series data obtained from the Helioseismic and Magnetic Imager (HMI; Scherrer et al. 2012) on board the SDO were preprocessed toward the force-free field condition (Wiegelmann et al. 2006), and were then used as the boundary conditions for the NLFFF extrapolation. The extrapolation was performed within a box of $840 \times 448 \times 448$ uniform grid points, corresponding to $300 \times 160 \times 160$ Mm (Jing et al. 2017).

### 3. Observations and Analysis

The main phase of the M6.5 flare (SOL2015-06-22T18:23) occurred in NOAA AR 12371, located at ($223''$, $183''$). In Figure 1(a), the light curve of the GOES SXR flux shows two episodes of small-magnitude emissions within the hour before the M6.5 flare. The onset times of the two precursors are 17:24 UT and 17:42 UT, which are denoted as P1 and P2, respectively. The $n$, $T$, and EM maps at P1 and P2 derived from the AIA data are shown in panels (b)–(g) of Figure 1. As mentioned earlier, the two precursors are well confined locally near the PIL (see Figure 2 for more details), enabling a direct comparison of different wavelengths. As shown in Figures 1(f)–(i), there is a spatial correlation among the AIA EUV, RHESSI HXR, and GST H$\alpha$ emissions.

Figure 2 further demonstrates their spatial correlation and confinement with the magnetic field. The top and bottom panels of Figure 2 show the emission maps of four wavelengths at the precursor times P1 and P2, respectively. A slit is set across the emissions, and the cross-sectional photometric intensity profiles of these emissions are plotted and compared with the magnetic field over the slit. Based on the comparison of the extrapolated magnetic fields and those derived from the MW spectrum fitting (Wang et al. 2017), it was found that the precursors in MW occurred in a strong magnetic field region (1200 G) around the flaring PIL. As shown in the middle panels of Figure 2, the emissions of AIA EUV, RHESSI HXR, and GST H$\alpha$ are also confined in similar local areas, while the SXR emission from the GOES 15 Solar X-ray Imager (SXI; Hill et al. 2005; Pizzo et al. 2005) extends to a substantially larger region.

Figure 4 shows the temporal variations of $T$, derived from four data sets using different methodologies, during the precursor phase. The RHESSI result is only partially shown for the first precursor, as the spectrum before 17:25 UT does not qualify for performing a reliable fitting, due to being at RHESSI night time. At a glance, these $T$ curves show a large discrepancy in the order of magnitude, but, as a general trend, they all reach their maxima around the two precursor times. Specifically, of the four instruments, EOVSA MW exhibits the highest temperature value and changes most rapidly during the precursor times, while GOES SXR and AIA EUV show the lowest temperatures, changing more gradually. The temperature of 15 MK is a clear line of demarcation between EOVSA and AIA/GOES. The range of 10–24 MK is where the temperature variation of RHESSI HXR is located. For the first precursor, the $T$ of EOVSA peaks the earliest, followed by the $T$ of AIA and GOES. The $T$ peak of RHESSI is unknown, due to the incomplete temporal coverage of the RHESSI fitting results. Such an order is not surprising, considering the empirical tendency for the HXR (or MW) emission to temporally coincide with the time derivative of the SXR emission of a solar flare, known as the Neupert effect (Veronig et al. 2008). For the second precursor, however, the $T$ peak of EOVSA lags behind that of RHESSI by $\sim$100 s. We plotted the time derivative of the GOES SXR light curve (the gray line in the top panel of Figure 1) and found that its peak coincides in time with the peak of the RHESSI HXR emission, as the Neupert effect indicates. The temporal delay between the MW and HXR emissions has been reported before in some events, but often on the order of a magnitude of seconds (Silva et al. 2000). The 100 s delay presented here is certainly not expected. The possible reasons are discussed in Section 4.

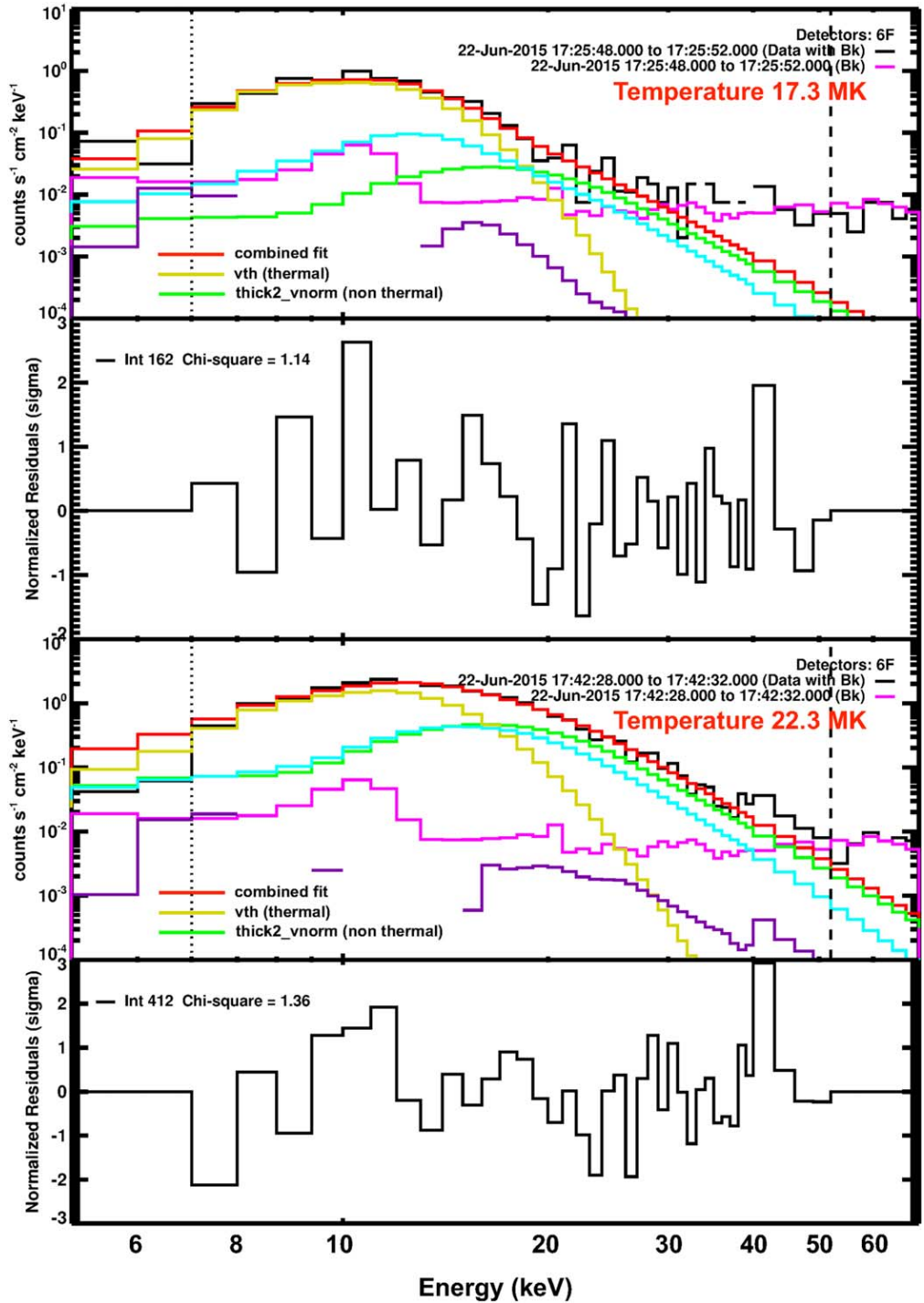

**Figure 3.** Spectral fitting results of RHESSI HXR in photon counts (black) and normalized residuals at two precursor peak times. The yellow, green, and red lines in each panel show the modeling results of Variable Thermal Modes (*vth*), Nonthermal Modes (*thick2_vnorm*), and the combined fitting results, respectively. The pink lines indicate the background values. The dashed lines mark the energy ranges during the fittings. The cyan and purple lines show the *albedo* and *pileup_mod* corrections, respectively.

Likewise, Figure 5 shows the temporal variation of EM, derived from AIA EUV, GOES SXR, and RHESSI HXR data, during the precursor phase. The peak EM of RHESSI HXR is at the same level as the peak EM of AIA EUV and GOES for the first precursor ($10^{48}$ cm$^{-3}$), at almost the same time. However, the second peak of AIA EUV is hard to distinguish, because EM is constantly increasing. Likewise, the second peak of RHESSI HXR is unknown, because of the data gap. There is a striking similarity between the EM curve obtained from the AIA EUV data and that from the GOES SXR data, especially during the period of the first precursor. Starting from $2 \times 10^{48}$ cm$^{-3}$, the EM profiles of AIA EUV and GOES SXR reach $5 \times 10^{48}$ cm$^{-3}$ during the precursor times. The EM curves obtained from the RHESSI HXR data, however, change more rapidly over a wide range ($1.5 \times 10^{47}$ cm$^{-3}$ to $3 \times 10^{48}$ cm$^{-3}$), but always less than AIA and GOES.

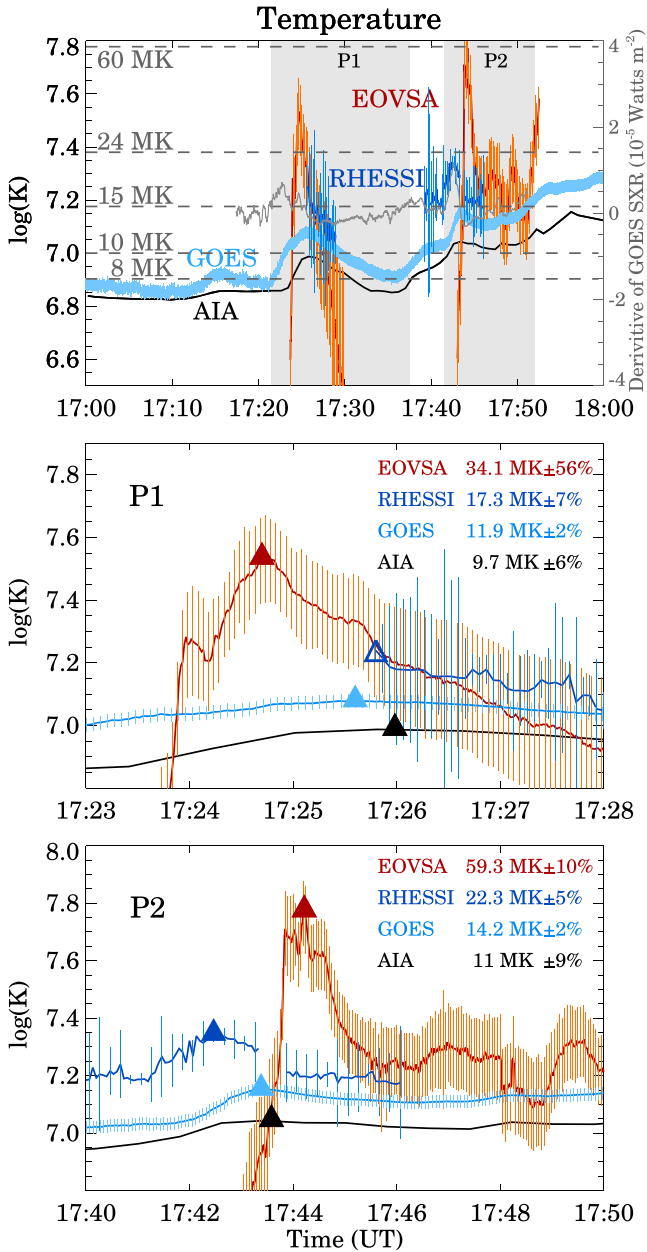

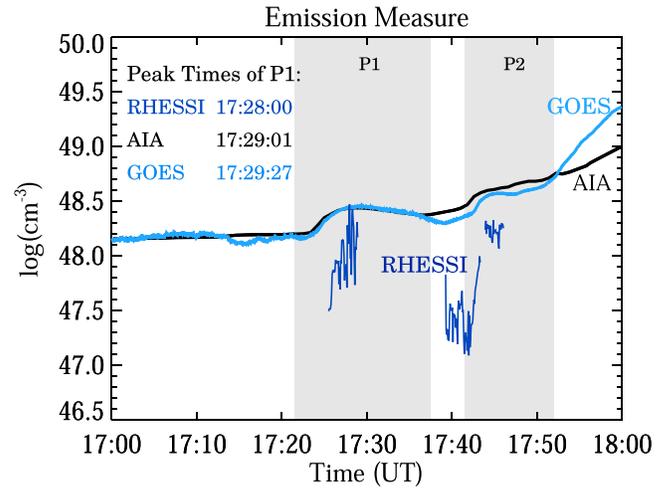

**Figure 5.** Temporal variations of EM (log cm$^{-3}$) and peak times of AIA (black), GOES (light blue), and RHESSI (dark blue) from 17:00 UT to 18:00 UT. The gray shaded areas P1 and P2 indicate the two precursor periods.

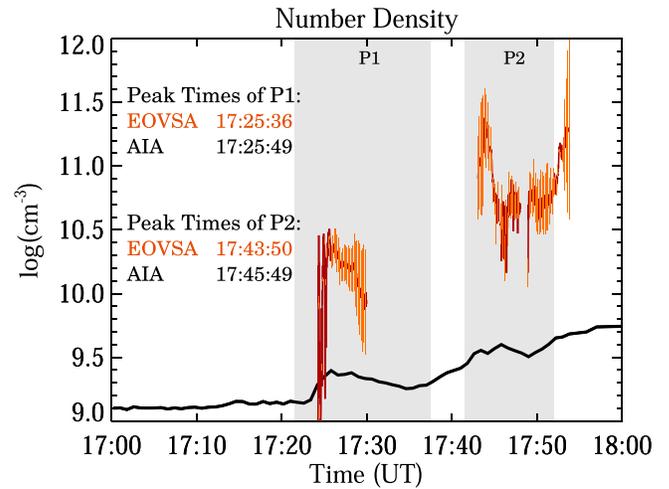

**Figure 6.** Temporal variations of number density ($n$) and peak times of AIA (black) and EOVSA (red) from 17:00 UT to 18:00 UT. The gray shaded areas P1 and P2 indicate the two precursor periods.

**Figure 4.** Top panel: temporal variations of temperature, derived from AIA (black), GOES (light blue), RHESSI (dark blue), EOVSA (red), and the derivative for GOES SXR (gray) during the precursor phase. The gray shaded areas P1 and P2 indicate the two precursor periods. Middle and bottom panels: magnified views of the temporal variations of $T$ during the two precursor periods. The solid triangles mark the peaks of the corresponding curves. Their values and uncertainties are listed in each panel. The hollow dark blue triangle in the middle panel indicates that the RHESSI peak of the first precursor is unknown because of the data gap before 17:25:00UT.

Figure 6 shows the temporal variation of $n$, derived from the AIA EUV and EOVSA MW data, during the precursor phase. Despite their significant differences in magnitude, both $n$ curves show peaks at the two precursor times, and the EOVSA MW emission always peaks ahead of the AIA EUV emission. It is not surprising, considering that the temporal variations of AIA EUV and GOES SXR are almost identical.

Figure 7 shows the time variations of AIA $\Delta$EMs integrated over different temperature ranges of log($T$) = 0.1. Concerning the two flare precursors, there are generally two types of curves in the figure: the ones with two clear peaks at the precursor times, and the ones without. For the former ones, the higher the temperature level, the clearer are the peaks observed during the precursor times, except for the black curve, which shows the lowest $\Delta$EM at the lowest $T$. It partially explains why AIA's $T$ peaks always come before the peaks of EM or $n$ (Figures 4–6). Curves with a temperature of log($T$) < 6.65 ($T$ < 4.5 MK) constitute another type. The magnitudes of $\Delta$EMs are greater at higher temperature levels, but they do not display clear changes during the precursor times.

Table 1 summarizes the numerical intervals of $T$, EM, and $n$ of the two precursors, derived from different data sets, as shown in Figures 4–6. As a response to the precursors, the values of these parameters vary, depending on the different instruments. Specifically, the temperatures as measured by AIA and GOES increase by ∼3 MK and ∼5 MK, respectively, during both of the two precursor times. The temperature measured by RHESSI increases by ∼10 MK, from 14 to 24 MK, for the second precursor, while the temperature measured by EOVSA increases even more, by more than 50% as that of RHESSI, from 25 to 59 MK. On the other hand, EM and $n$ as measured by RHESSI and EOVSA, respectively, increase by at least 10 times during both of the two precursor phases, while these two parameters as measured by AIA and GOES increase only by the same order of magnitude. In general, the measurements by AIA and GOES show similar lowest numerical values and the fewest variations, even though their results were obtained independently by two different methodologies. To conclude, the increases of the thermal parameters detected by AIA and GOES are at the same level. EOVSA's thermal parameter changes at least 10 times as much as that of AIA in the measurements of $n$ and $T$. As for the HXR emission, the temperature increase detected by RHESSI is at least twice as much as that of AIA and GOES.

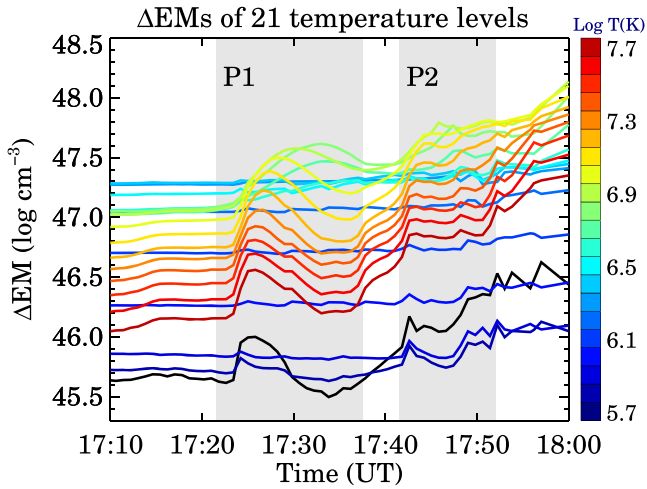

**Figure 7.** Time profiles of AIA ΔEMs integrated over 21 temperature ranges of log(T) = 0.1. Each curve shows the temporal variation of the differential EM at the corresponding temperature level. High-temperature levels (log (T) > 6.65) show clear peaks at two precursor times. Low-temperature levels (log(T) < 6.65) do not have clear peaks, except for the curve with the lowest temperature (log(T) = 5.65).

## 4. Conclusion and Discussion

To summarize, we present a case study of the temporal variation of $T$, $n$, and EM derived from different data sets during a flare precursor period. The results from different data sets show apparent temporal consistency among the thermal parameters in multiwavelengths, as well as significant quantitative differences, which is likely due to the different emission mechanisms as well as the different methodologies applied in the data analysis of the different instruments. During the precursor phase, the temperature as measured by AIA/GOES, RHESSI, and EOVSA varies over the ranges of 8–15 MK, 10–24 MK, and 15–60 MK, respectively.

Of all the available measurements, RHESSI has the smallest EM value, which varies from $1.5 \times 10^{47}$ cm$^{-3}$ to $3 \times 10^{48}$ cm$^{-3}$, and EOVSA has the largest number density variation, from 1 to $3 \times 10^{10}$ cm$^{-3}$. AIA/GOES has the most gentle variations in EM, of 1.5–4 × 10$^{48}$ cm$^{-3}$, and $n$, of 2–4 × 10$^9$ cm$^{-3}$.

Note that EM and $n$ have the following relationship:

$$EM = \int n^2 dV. \quad (7)$$

Our results are summarized as follows:

1. GOES SXR and AIA EUV have almost identical EM variations (especially before 17:52 UT), and very similar $T$ variations (especially after 17:23 UT). During the precursor phase, both EM and $T$ as measured by the GOES SXR and AIA EUV passbands are raised to twice their initial values ($T$ increases from 8 to 15 MK, and EM increases from 1.5 to $3 \times 10^{48}$ cm$^{-3}$).
2. Compared to GOES SXR and AIA EUV, RHESSI HXR shows greater temperature changes at the 15 MK level and above. EM as measured by RHESSI HXR during the precursor phase is 10 times higher than it was before the precursors. For the first precursor, RHESSI HXR, GOES SXR, and AIA EUV have very close peaks, no matter the magnitude or temporal sequence.
3. The $T$ peak as measured by EOVSA MW (59 MK) is almost three times higher than the $T$ peak as measured by AIA EUV and GOES SXR (15 MK). The $n$ peak of EOVSA MW ($3 \times 10^{10}$ cm$^{-3}$) is more than 10 times higher than the $n$ peak of AIA EUV ($3 \times 10^9$ cm$^{-3}$). EOVSA MW exhibits high thermal variations of $T$ and $n$, and it has the greatest uncertainties in its measurements as well (Table 1).

It is clear that AIA and GOES, compared to RHESSI and EOVSA, show lower temperatures and smoother variations. Such a result is not very surprising, as different instruments, operating at different wavelengths, inherently are sensitive to different temperature ranges. Moreover, the difference is also a result of the different emission mechanisms at play as well as a result of different emitting area selections being used in the temperature calculation. For AIA and GOES, the temperature measurements are averaged over a large area. Specifically, the AIA temperature is averaged within the black boxes in Figures 1(b) and (d), and GOES receives emissions from the whole solar disk. On the other hand, the temperature derived from the EOVSA MW data reflects the instantaneous thermal behavior within a small area of the precursor brightening, and the temperature derived from the RHESSI HXR data is calculated within the footpoints and loop top of the HXR emission.

The variations of temperature for AIA and GOES are in good agreement with each other. However, there is a constant difference of about 3–5 MK in their magnitudes. As shown in Figure 7, the hot components (>4.5 MK) of AIA DEM increase from 17:13 UT, whereas the cold components (<4.5 MK) remain unchanged. Considering that AIA EUV and GOES SXR have almost identical EM variations (Figure 5), the difference in temperature between AIA EUV and GOES SXR is probably due to the unchanging cold components of the AIA emissions. This also explains why the difference is not so obvious until 17:13 UT, which is when the hot components start to increase.

The temperature derived from the EOVSA MW data shows a larger variation and a higher maximum value than that derived from the RHESSI HXR data. This is probably because the emitting area used in calculating the MW temperature is much smaller than that used in calculating the RHESSI temperature, i.e., $10'' \times 30''$ versus $\sim 30'' \times 50''$, according to the HXR images. Besides the difference in magnitudes, we notice a significant time delay (100 s) between the temperature peak of RHESSI HXR and that of EOVSA MW, observed from the second precursor. The time delay of the MW flux peaks relative to the HXR flux peaks has been known for a long time. For

Table 1
Data Source List of Parameters

|  | log(T) unit K | log(n) unit cm$^{-3}$ | log(EM) unit cm$^{-3}$ |
|---|---|---|---|
| AIA | 6.75 ($\pm 6\%$) – 7.05 ($\pm 9\%$) | 9.20 ($\pm 57\%$) – 9.60 ($\pm 68\%$) | 48.20 ($\pm 1\%$) – 48.60 ($\pm 2\%$) |
| GOES | 6.90 ($\pm 2\%$) – 7.15 ($\pm 2\%$) | N/A | 48.15 ($\pm 5\%$) – 48.58 ($\pm 5\%$) |
| RHESSI | 7.05 ($\pm 40\%$) – 7.35 ($\pm 5\%$) | N/A | 47.09 ($\pm 63\%$) – 48.47 ($\pm 27\%$) |
| EOVSA | 6.50 ($\pm 56\%$) – 7.79 ($\pm 10\%$) | 9.85 ($\pm 81\%$) – 11.4 ($\pm 31\%$) | N/A |

**Note.** The lower limits of T from EOVSA and RHESSI are selected at the end of the first precursor (∼17:30 UT), due to the data gap before the first precursor peak. For AIA and GOES, the lower limits are the minimum values before the first precursor. All upper limits are selected at the peak times of the second precursor (∼17:42 UT). 'N/A' means that this parameter was not obtained from the corresponding method.

example, a statistical study of 57 bursts from 27 solar flares shows that such delays are $6 \pm 5$ s for impulsive flares and $15 \pm 6$ s for nonimpulsive ones (Silva et al. 2000), which are, however, much shorter than the time delay presently under discussion. Several ideas have been adopted for interpreting time delays: the delay is either due to the trapping effect of nonthermal electrons in the loop top and the energy dependence of Coulomb collisions (e.g., Lee & Gary 2000), or it is due to other generic loss mechanisms (Kundu et al. 2001; Lee et al. 2002). The trapping of nonthermal electrons in the loop top may arise due to magnetic mirroring and the energy-dependent Coulomb collisions, because it is more difficult for higher-energy electrons to be scattered into the loss cone than lower-energy electrons (Lee & Gary 2000). This scenario can explain the observed time delay, if the high-energy (>300 keV) electrons are responsible for the MW emissions and the low-energy (20–200 keV) electrons for the HXR emissions. However, the time delay (100 s) that we found for the second precursor is unusually long. In a more general approach, Lee (2005) suggested that the trapping effect can be severe for strongly converging magnetic fields and extended electron ejection times, in which case our observation may possibly be explained. We anticipate that the present observation would motivate theoretical modeling of magnetic evolution combined with the participation of the thermal process in the future.

We appreciate the referee for detailed comments that improved this paper significantly. We thank the NASA SDO team for the AIA data. AIA is an instrument on board SDO, NASA's mission under the Living With a Star (LWS) program. We thank the EOVSA team for MW data and the RHESSI team for HXR data. We are grateful to Dr. Mark Cheung for his support on AIA DEM analysis, Dr. Gregory D. Fleishman for his efforts in processing the EOVSA MW spectrum data, and Dr. Jeongwoo Lee for his constructive comments on the manuscript. Special thanks to Dr. Brian R. Dennis for his professional advice on the RHESSI HXR fitting. This work is supported by NSF under grants AGS-1927578, AGS-1954737, and AGS-1821294, and NASA under grants 80NSSC17K0016, 80NSSC18K0673, 80NSSC18K1705, 80NSSC19K0257, 80NSSC19K0859, 80NSSC21K0003, and 80NSSC21K1671.

## ORCID iDs

Ju Jing https://orcid.org/0000-0002-8179-3625
Haimin Wang https://orcid.org/0000-0002-5233-565X